\documentclass[11pt,paper,twocolumn]{article}
\usepackage[square,sort,comma,numbers]{natbib}
\usepackage[utf8]{inputenc}
\usepackage{amsmath}
\usepackage{amsfonts}
\usepackage{amssymb}
\usepackage[cm]{fullpage}
\usepackage{soul}
\usepackage{pifont}
\usepackage{wrapfig}
\usepackage{xcolor}
\usepackage{xfrac}
\usepackage{multicol}
\usepackage{subcaption}

\definecolor{green}{HTML}{008000}
\definecolor{red}{HTML}{800000}
\definecolor{blue}{HTML}{1C789E}
\definecolor{chocolate}{rgb}{0.82, 0.41, 0.12}

\newcommand{\si}{\sigma}

\newcommand{\varep}{\varepsilon}
\newcommand{\ig}{\includegraphics}
\newcommand{\lw}{\linewidth}
\newcommand{\bi}{\begin{itemize}}
\newcommand{\ei}{\end{itemize}}
\newcommand{\bra}{\langle}
\newcommand{\ket}{\rangle}
\newcommand{\bc}{\begin{center}}
\newcommand{\ec}{\end{center}}

\newcommand{\tchoc}[1]{\textbf{\textcolor{chocolate}{#1}}}
\newcommand{\pref}[1]{(\ref{#1})}
\newcommand{\fref}[1]{Fig. \ref{#1}}
\newcommand{\beq}{\begin{equation}}

\newcommand{\eeq}{\end{equation}}
\newcommand{\bfgr}{\begin{figure}}
\newcommand{\efgr}{\end{figure}}

\usepackage{wasysym}
\usepackage{hyperref}
\usepackage{mathtools}
\usepackage{caption}

\DeclarePairedDelimiter\rounding{\lfloor}{\rceil}
\usepackage{authblk}
\usepackage{titling}
\usepackage{graphicx}	
\usepackage[export]{adjustbox}
\usepackage{natbib}

\date{}
\author{Victor Herdeiro}
\affil{Department of Mathematics, King’s College, London, United Kingdom.}
\title{\textbf{Numerical estimation of structure constants in the 3d Ising CFT through Markov chain UV sampler}}
\begin{document}
\twocolumn[{%
 \begin{@twocolumnfalse}
 
\maketitle
\thispagestyle{empty}

\begin{abstract}
\href{https://journals.aps.org/pre/abstract/10.1103/PhysRevE.94.043322}{[Herdeiro\;\&\;Doyon Phys.\,Rev.\,E (2016)]} introduced a numerical recipe,  dubbed UV sampler, offering precise estimations of the CFT data of the planar two-dimensional critical Ising model. It made use of scale invariance emerging at the critical point in order to sample finite sublattice marginals of the infinite plane Gibbs measure of the model by producing ``holographic" boundary distributions. The main ingredient of the Markov chain MonteCarlo (MCMC) sampler is the invariance under dilation. This article presents a generalization to higher dimensions with the critical 3d Ising model. This leads to numerical estimations of a subset of the CFT data - scaling weights and structure constants - through fitting of measured correlation functions. The results are shown to agree with the recent most precise estimations from numerical bootstrap methods \href{https://arxiv.org/abs/1603.04436}{[Kos,\;Poland,\;Vicci\;\&\;Simmons-Duffin JHEP (2016)]}.
\end{abstract}
\vspace{1cm}
\end{@twocolumnfalse}
}]


%


\section*{Introduction}

\paragraph{The 3d Ising model.} The Ising model is a milestone of statistical physics. It consists in a statistical model on a graph with binary random variables and nearest-neighbour interactions. Having been studied for more than a century now, it has given a lot of insight in the fields of materials physics and critical phenomena, amongst others.

One of its strength lies in its easiness to be generalized to any dimensions or even fractal graphs. Peierls's argument \citep{peierls1936ising} brought a satisfying qualitative proof of the existence of a critical point for the two dimensional case. The generalization of his droplets to higher dimensions implied that such ordered-disordered phase transition had to exist for any dimension $d\geq 2$. For $d=2$, the model has since been solved exactly by Onsager \citep{onsager1944crystal}; while for $d\geq 4$ it has been proven that the Landau–Ginzburg theory gives the exact value of the critical exponents \citep{landau1950theory}. Only $d=3$ has resisted every attempt at an exhaustive solution so far.

Recent breakthroughs by means of the conformal bootstrap program applied to 3d CFTs, e.g. the constraints of associativity and positivity on the CFT operator algebra, have led to the most precise estimation of the critical exponents and some structure constants \citep{el2012solving,kos2016precision,simmons2016tasi,poland2016conformal}. The numerical precision of this approach have far overtaken previous MonteCarlo or analytical expansion results.

The goal of this paper is to extend the numerical procedure of \citep{herdeiro2016monte} to the 3d Ising model. This numerical procedure, dubbed UV sampler, is a MonteCarlo Markov chain (MCMC) allowing to sample sublattice marginals of infinite-volume statistical models at criticality, by effectively producing a ``holographic" boundary condition that encodes the infinite volume beyond it. It thus gives access directly to bulk data of the critical point, without the need for finite-size scaling. Catching up on the high precision of bootstrap method seems out of reach of the numerical procedure introduced here thus the aim is more of presenting an alternative approach using this MonteCarlo Markov chain and free of the bootstrap's assumptions. Its results will be shown to agree with the state of the art knowledge on the 3d Ising universality class, giving a precision improvement of the MonteCarlo estimations of the structure constants ($C_{\varep \si \si}$, $C_{\varep \varep \varep}$) in this model.

\paragraph{Markov chain holographic sampling.} Ref. \citep{herdeiro2016monte} studied the planar critical 2d Ising model through a MCMC.  It showed that an implementation of dilations on a lattice statistical model at criticality coupled with a sufficiently long rethermalization step through fixed boundaries Swendsen-Wang (SW)  lattice flip updates \citep{swendsen1987nonuniversal} would eventually mix into a Markov chain sampling the distribution of the sublattice marginal in the infinite plane Gibbs measure. Ref. \citep{herdeiro2017markov} showed a successful generalization of this method and its results to the $O(1<n\leq 2)$ loop gas models where non-local contributions have to be accounted for.

Let us recall the main arguments of \citep{herdeiro2016monte}. Working on the complex plane with a radially quantized generic (Euclidean) CFT and picking $A$ a disk centered on the origin, the subdomain marginal $\Psi_{\partial A}$ is defined by
\beq
\label{eq:marginalDistribution}
\Psi_{\partial A}  = \int {\cal D} \phi_{\mathbb{C} \setminus A}\ e^{-S_{\mathbb{C} \setminus A}[\phi]}.
\eeq
It can be enhanced to include operator insertions in $\mathbb{C} \setminus A$. Knowledge of $\Psi_{\partial A}$ allows to measure any infinite plane correlators with any insertions ${\cal O}_1(x_1){\cal O}_2(x_2)\dots{\cal O}_n(x_n)$ as long as $x_1,x_2,\dots,x_n \in A$. In this sense it is the \textit{bulk marginal} of $A$. An interesting observation is that the information on the marginal distribution is entirely contained on $\partial A$ because of ultra-locality of the measure. In this respect it is ``holographic": it is a probability distribution on the set of boundary states of $A$, which reproduces, from the viewpoint of observables in $A$, the statistical information outside $A$. The goal of the UV sampler is to approximate this marginal distribution by a \textit{typical} sample of $\Psi_{\partial A}$ - \textit{typical} in the sense of Markov chain sampling.

We introduce the $R_{\lambda}\,[\cdot]$ operation which acts as $$R_{\lambda}:\quad \Psi_{\partial A} \to \lambda \bullet \Psi_{\lambda^{-1} \partial A}\ .$$ It maps a state on $\lambda^{-1}\partial A$ to a state on $\partial A$ by use of a dilation of parameter $\lambda > 1$. For a Gibbs measure endowed with scale invariance, it is shown in \citep[II. A.]{herdeiro2016monte} that the fixed point of such transformation is a MCMC sampling the infinite plane marginal on $A$. The argument is broadly as follows:

\bi
\item[$\cdot$] We start with any boundary condition $\Psi^{\rm gen}_{\partial A}$ (where ``gen'' stands for generic).
\item[$\cdot$] We construct a chain of iterations of $R_{\lambda}$ where $\lambda > 1$ (in fact, it does not need to be the same on every link): $$ \Psi^{\rm gen}_{\partial A} \xrightarrow[]{R_{\lambda}} R_{\lambda} [ \Psi^{\rm gen}_{\partial A} ] \xrightarrow[]{R_{\lambda}} \dots \xrightarrow[]{R_{\lambda}} R_{\lambda}^n [ \Psi^{\rm gen}_{\partial A}] \xrightarrow[]{R_{\lambda}} \dots$$

This chain converges to its fixed point which must be a marginal invariant under rescalings. Ref. \citep[II. A.]{herdeiro2016monte} argues that the limit is the distribution obtained when integrating out all fields outside $A$, by (\ref{eq:marginalDistribution}):
$$ R^n_{\lambda} [ \Psi^{\rm gen}_{\partial A}] \to \Psi_{\partial A}\quad ( n \to \infty)$$
\ei
Scale invariance is an essential tool in constructing the marginal distribution. 


Using this ingredient, results in \citep{herdeiro2016monte} evidenced that finite size and boundary effects could be reduced to negligible levels. This method was dubbed UV sampler as it can be seen as a succession of inverse Kadanoff transformations or a RG flow towards a UV fixed point \citep{massiveQFT}. 


\section*{Ising 3d UV sampler}

\paragraph{Implemented discrete lattice dilations.}

We define $A$ as a finite volume connected subset of $\mathbb{R}^3$, and $L_{A} = A \cap \mathbb{Z}^3$ the (square) sublattice contained within. In the implementation choice detailed hereafter, $A$ is a simple cubic box. When applying a dilation of parameter $\lambda>1$, mapping $\lambda^{-1} A \to A$ induces a dilation on the sublattice $L_{\lambda^{-1} A} \to L_A$ such that, for\footnote{bold variables stand for lattice vectors} $\si^L_{\bf i} \in \{-1,1\}\ \forall\, {\bf i} \in L_A$ the binary lattice variables of the Ising model are mapped as
\begin{equation}
\label{eq:latticetransformation}
\si^L_{\rounding{\lambda^{-1} {\bf i}}} \to \si^L_{\bf i},
\end{equation}
where $\rounding{\cdot}$ returns the closest site on $L_{\lambda^{-1} A}$. From this point of view, the dilation is not a transformation on $L_A$  in the sense that the lattice spacing is not dilated but a transformation on the values taken by $\si^L$ on $L_A$ - i.e. on a configuration $\bar{\si} = \{\si^L_{\bf i}:\;{\bf i}\in L_A\}$. It is a basic fact of discrete systems that there is more information on $L_A$ than on $L_{\lambda^{-1} A}$ meaning that (\ref{eq:latticetransformation}) is a one-to-many mapping.

If applying straightforwardly this mapping, a value $\si_{\bf i}^L$ with ${\bf i} \in \lambda^{-1}A$ could be mapped to many images in $A$. This means that a `pixel' or lump of many identically oriented neighbour sites will be formed. This corresponds to an excess of short distance correlations. It is dubbed a `pixel' dilation.

In \citep{herdeiro2016monte}, another prescription was used. First $L_A$ is partitioned by the equivalence class ${\bf i} \sim {\bf j}$ iff $\rounding{\lambda^{-1} {\bf i}} = \rounding{\lambda^{-1} {\bf j}}$. Second, within each subset of the partition a single site $\bf i$ is randomly picked and (\ref{eq:latticetransformation}) is performed on it. This forces (\ref{eq:latticetransformation}) to be applied in a one-to-one fashion. Last, every unpicked site is unassigned at this stage, in other words left as a `hole'; a heat-bath procedure is applied on each one of them to assign a spin value depending solely on its first neighbours. Since first neighbours can be missing - either because of being holes themselves, or because they are not in $A$ - this choice induces a lack of smallest distances ($\si_{\bf i} \si_{\bf i+1}$) correlations. This is dubbed `heat-bath' dilation.

It must be reminded that the re-thermalization steps that come after each dilations are performed with fixed boundaries, fixed $\si_{\bf i}$ for ${\bf i} \in \partial L_A$. This has the effect of propagating the new boundary condition inside $L_A$. The effects of any dilation prescription only remain on and near the boundary, where the sites will not be updated after discrete dilation.

As a general rule when going 2d $\to$ 3d, a larger fraction of the spins are close to the boundary. The direct implication is that the prescription which was good enough in 2d - in \citep{herdeiro2016monte} - could leave too strong boundary effects for precision fitting in 3d. Anticipating such issues, an hybrid of the pixel and heat-bath prescriptions was implemented. It follows the same partition and second step as heat-bath but in the last stage, on each empty lattice site, at random and with equal probability, either \pref{eq:latticetransformation} is applied or heat-bath assignment is used. This is named `hybrid' discrete lattice dilations and has been used in every case throughout this numerical work.

\paragraph{MCMC description.}

The starting point of the MCMC is a 200x200x200 cubic Ising lattice in a vaccum ordered state. At zero external magnetic field and normalizing the nearest-neighbour interaction strength $J$ to 1, the unique coupling is the inverse temperature $\beta$ which is set to the best known estimation of its critical value $\beta_c$ \citep{talapov1996magnetization}: $$\beta = \beta_c \approx  0.221\,654\,4\ (3).$$

The following steps are taken for the MCMC to produce samples of holographic boundaries:

\begin{enumerate}
\item \textit{Torus intermediate state:} 200 lattice flips (Swendsen- Wang (SW) flips \citep{swendsen1987nonuniversal}) are applied, with periodic boundary conditions. The goal here is to bring the vaccum state to a torus critical state, which is closer to the critical plane, with relatively small CPU effort.
\item \textit{Dilations and rethermalizations:} The chain enters then the cycle of hybrid discrete lattice dilations followed by fixed-boundaries SW flips. The parameters - the scaling factor $\lambda$ and the number $N_{\rm SW}$ number of lattice updates - take successive values as per following lists:

\begin{align*}
\{\lambda\} = &\underbrace{2,\ldots,2}_{5\ {\rm times}},\underbrace{1.5,\ldots,1.5}_{4},\underbrace{1.3,\ldots,1.3}_{4},\underbrace{1.2,\ldots,1.2}_{4},\\
&\underbrace{1.1,\ldots,1.1}_{4}, \underbrace{1.05,\ldots,1.05}_{7},\underbrace{1.03,\ldots,1.03}_{7}
\end{align*}

\begin{align*}
\{N_{\rm SW}\} = &\underbrace{150,\ldots,150}_{5},\underbrace{100,\ldots,100}_{16},\underbrace{80,\ldots,80}_{14}
\end{align*}

This means that the first pass is a $\lambda=2$ dilation followed by 150 SW flips, this being repeated for the next 4 passes, etc., until the last pass consisting in a $\lambda=1.03$ dilation followed by 80 lattice flips. This totals 35 passes of dilations and lattice flips. At first this can seem like a long process. These somewhat arbitrary numbers give a correctly mixing chain as the following results will show, but this implementation does not have any pretension of using the optimal parameters. They were found by trial and error. It takes $\sim$20 minutes\footnote{on a single core of a AMD Opteron$^{\rm tm}$ 6174 CPU} to complete this mixing step.

\item \textit{Measurements:} this Markov chain measures spatial correlations on the lattice. These correlations involve operators defined on the lattice: the spin lattice operator $\si^L_{\bf i}$ and the energy density operator $$\varep^L_{\bf i} \equiv \si^L_{\bf i} \si^L_{\bf i+1}.$$ The list of measured correlators is explicitly $$\bra \varep^L_{\bf i}\ket,\ \bra \si^L_{\bf i} \si^L_{\bf j} \ket,\ \bra \varep^L_{\bf i} \varep^L_{\bf j} \ket,\ \bra \varep^L_{\bf i} \si^L_{\bf j} \si^L_{\bf k} \ket\ {\rm and}\ \bra \varep^L_{\bf i} \varep^L_{\bf j} \varep^L_{\bf k} \ket.$$The long distance behaviour of such correlators is known from their expansion on the basis of the CFT operators:
\begin{align*}
\si^L_{\bf i} &= a^{\Delta_{\si}} N_{\si}\,\si(a{\bf i}) + \ldots\ ,\\
\varep^L_{\bf i} &= \bra \varep^L \ket I + a^{\Delta_{\varep}} N_{\varep}\,\varep(a{\bf i}) + \ldots\ ,
\end{align*}
where $a$ is the lattice spacing, setting $a=1$ from here on. This choice of correlators was picked on the principle that they would be enough to fit an interesting subset of the CFT data of the 3d Ising model, namely the scaling dimensions $\Delta_{\si}$ and $\Delta_{\varep}$, and the three-point couplings $C_{\varep \si \si}$ and $C_{\varep \varep \varep}$.

Some prescriptions were taken for the insertions choice ${\bf i}$, ${\bf j}$ and ${\bf k}$: no insertions were made less than $30$ lattice units afar from the boundaries - to remove possible remnant boundary effects - and ${\bf i}-{\bf j}$, ${\bf i} - {\bf k}$ and ${\bf j} - {\bf k}$ are always taken along the three lattice directions for computational simplicity. On top of this, for the last two correlators listed above, insertions were made with ${\bf j}$ and ${\bf k}$ diametrically opposed when sitting on ${\bf i}$, e.g. ${\bf j} -{\bf i} = -( {\bf{k}} - {\bf{i}} )$.

Each measurement is spaced by 5 SW flips. Every 20 measurements a new $\lambda = 1.03$ lattice dilation is applied followed by a rethermalization consisting in 80 SW flips. This aims at updating the boundary configuration and at diminishing the autocorrelations of the measurements \citep{herdeiro2016selfaveraging}.


\end{enumerate}

\paragraph{Two-point functions and correlators.}

\begin{figure*}[b]

\begin{subfigure}{0.48\linewidth}
\bc
\ig[width=.95\lw]{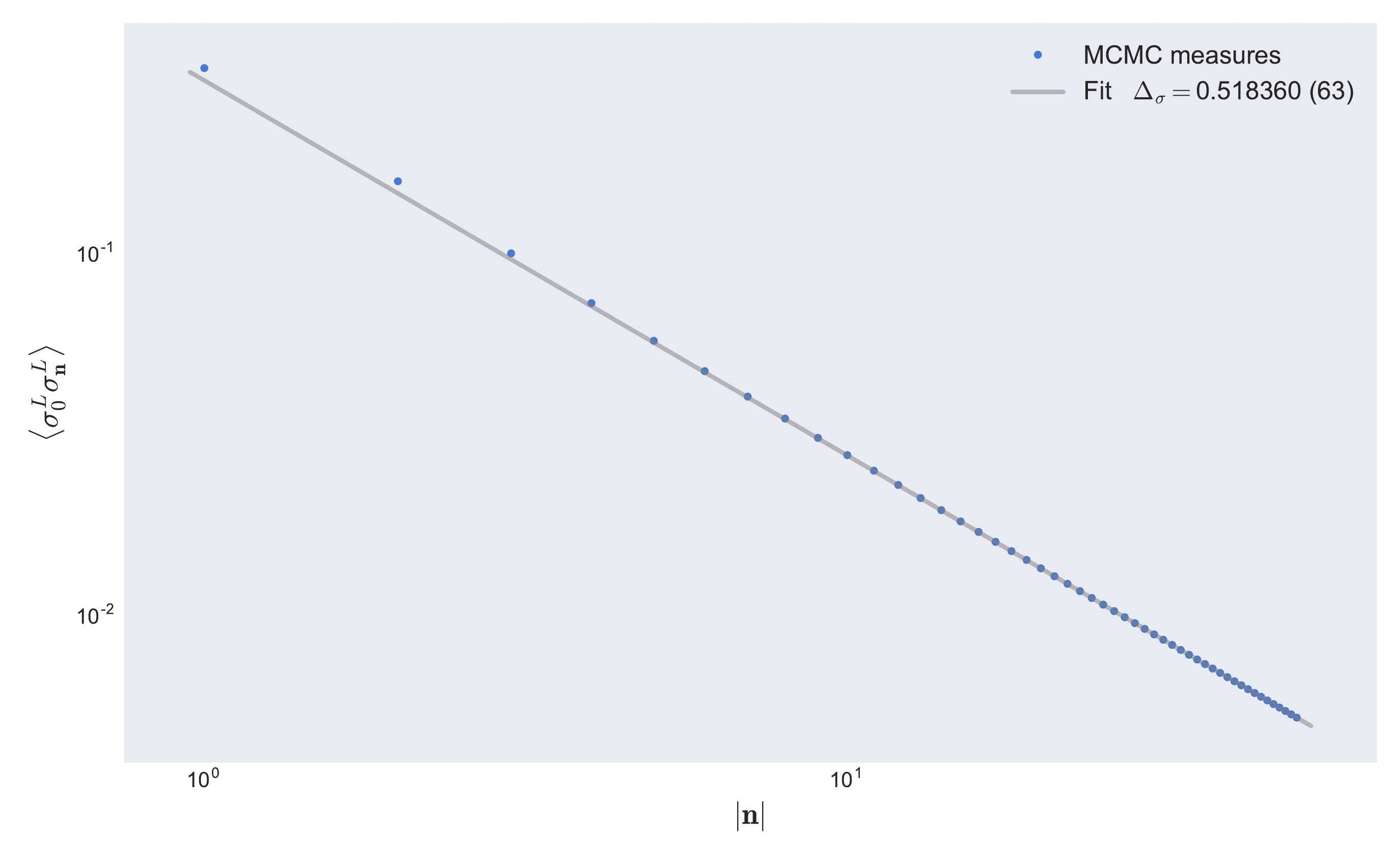}
\caption{\small{Graph of the $\bra \si^L_{\bf 0} \si^L_{\bf n} \ket$ correlations as a function of the separation $|{\bf n}| \in [1,50]$. The power law behaviour seems manifest. The gray line represents the most accurate power law fit. This is solid proof that the sample is very close to the planar expectation. 
}}
\label{fig:graphSS}
\ec
\end{subfigure}
\hfill
\begin{subfigure}{0.48\linewidth}
\bc
\ig[width=.95\lw]{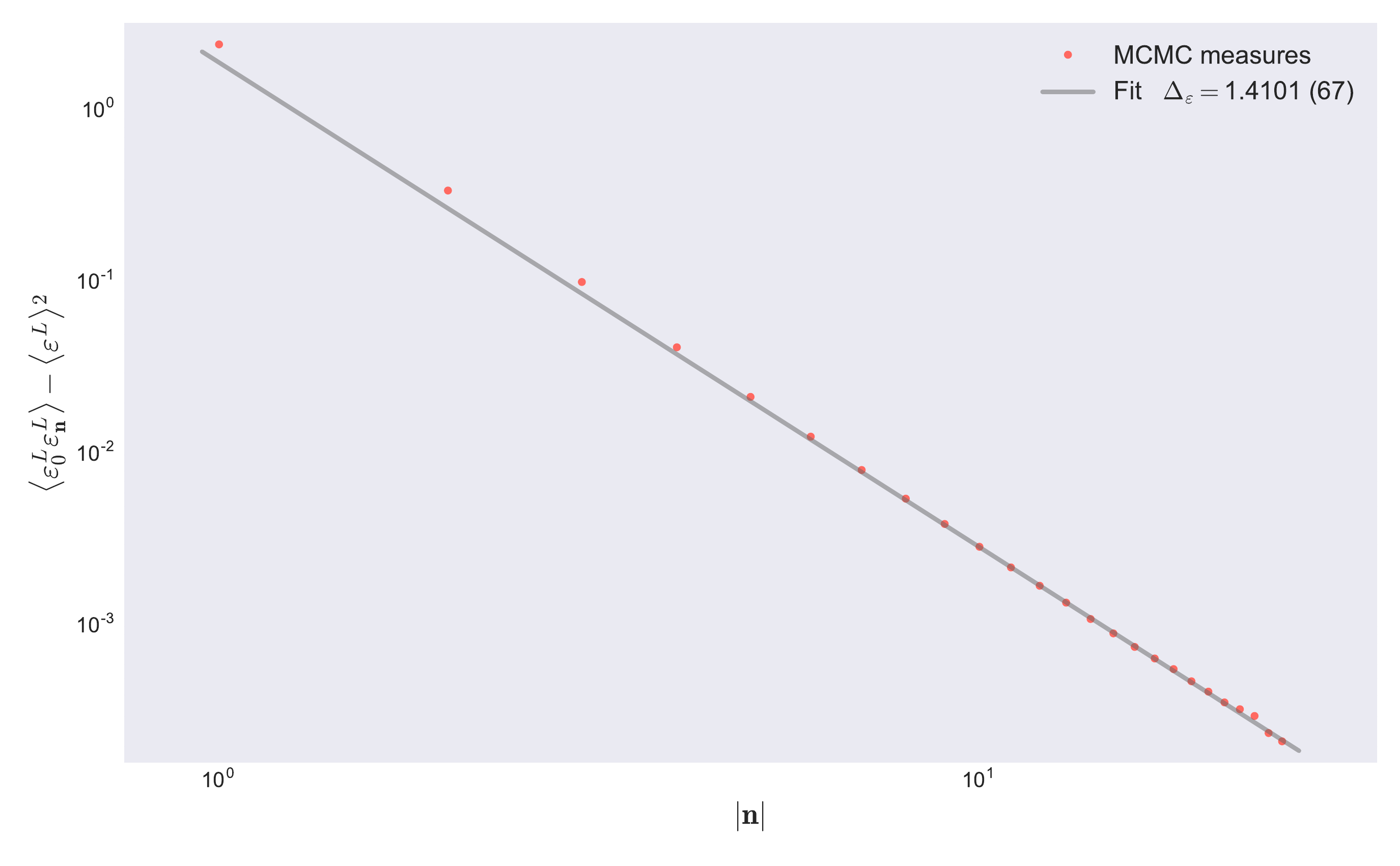}
\caption{\small{Graph of the $\bra \varep^L_{\bf 0} \varep^L_{\bf n} \ket - \bra \varep^L \ket^2$ connected two point function for separations $|{\bf n}| \in [1,25]$. As much as for $\si^L \si^L$ correlations, the power law behaviour is explicit. }}
\label{fig:graphEE}
\ec
\end{subfigure}

\begin{subfigure}{0.48\linewidth}
\bc
\ig[width=.95\lw]{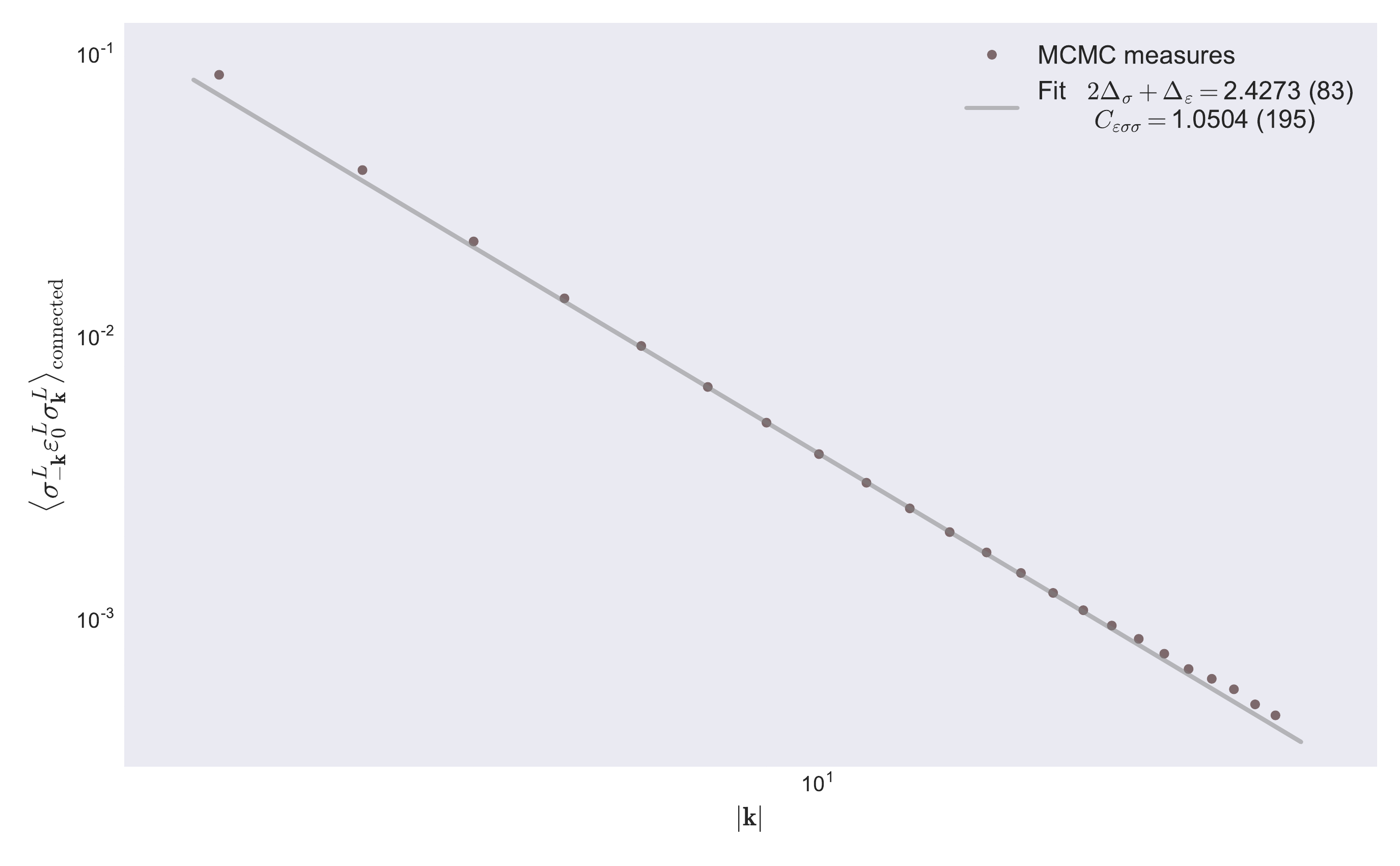}
\captionsetup{singlelinecheck=off}
\caption[]{\small{Graph of the connected part $\bra \si_{-\bf k}^L \varep_{\bf 0}^L \si_{\bf k}^L \ket - \bra \varep^L \ket \bra \si^L_{-\bf k} \si^L_{\bf k} \ket$  as a function of  $|{\bf k}|\in [3,25]$.  With the two $\sigma^L$ insertions diametrically opposed when sitting on the $\varep^L$ insertion point, CFT predicts a $\propto C_{\varep \si \si} |{\bf k}|^{-2\Delta_{\si} - \Delta_{\varep}}$ profile.}}

\label{fig:graphESS}
\ec
\end{subfigure}
\hfill
\begin{subfigure}{0.48\linewidth}
\bc
\ig[width=.95\lw]{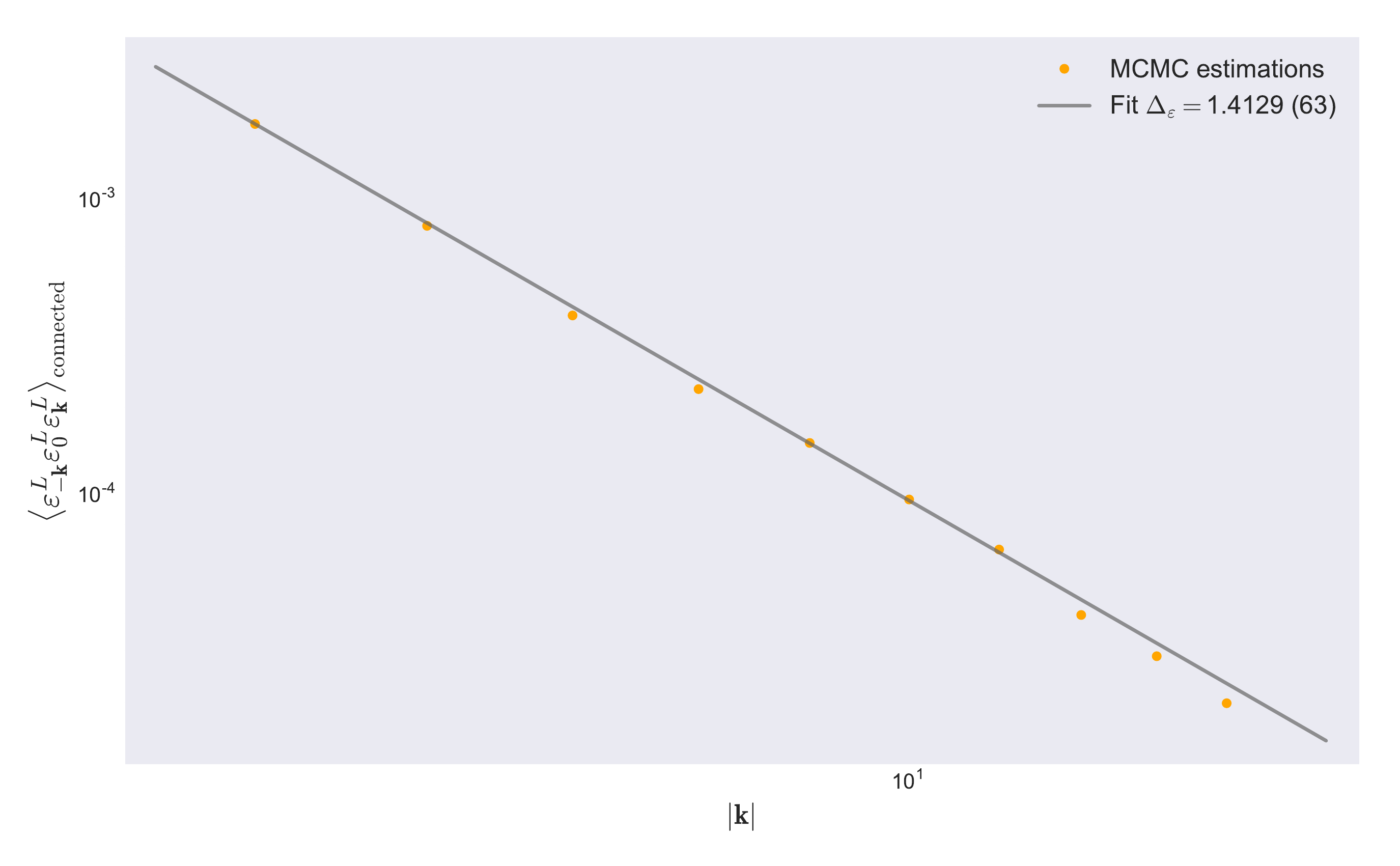}
\captionsetup{singlelinecheck=off}
\caption[]{\small{Graph of the connected part  $\bra \varep_{- \bf k}^L \varep_{\bf 0}^L \varep_{\bf k}^L \ket - \bra \varep^L \ket \big( \bra \varep^L_{-\bf k} \varep^L_{\bf 0} \ket + \bra \varep^L_{-\bf k} \varep^L_{\bf 0} \ket + \bra \varep^L_{\bf 0} \varep^L_{\bf k} \ket \big) - \bra \varep^L \ket^3$ as a function of  $|{\bf k}|\in [5,15]$. Conformal symmetry constrains the decay to be $C_{\varep \varep \varep}\, N_{\varep}^3\, 2^{-\Delta_{\varep}}  |{\bf k}|^{-3\Delta_{\varep}}$. }}
\label{fig:graphEEE}
\ec
\end{subfigure}

\caption{Measured lattice correlators}
\end{figure*}

Figures \ref{fig:graphSS}, \ref{fig:graphEE}, \ref{fig:graphESS} and \ref{fig:graphEEE} show the results collected for correlators $\bra \si^L \si^L \ket$, $\bra \varep^L \varep^L \ket$, $\bra \varep^L \si^L \si^L \ket$ and $\bra \varep^L \varep^L \varep^L \ket$ respectively. The number of samples of $\Psi_{\partial A}$ used is $9\,000$ for the first two graphs, $8\,000$ for the third one and $60\,000$ for the last one. For each graph, the axes are log-scaled and possible disconnected contributions - from the one-point expectation value of $\varep^L$ - were subtracted.

The fitted non universal quantities are contained in table (\ref{tab:offsets}), while universal quantities are splitted between tables (\ref{tab:weights}) and (\ref{tab:structureConstants}).

Comments specific to each observable are as follows:
\bi
\item $\bra \si^L \si^L \ket$, \fref{fig:graphSS}: Smallest separations, $|{\bf n}| \leq 5$, show the most departure from the fitted power law, this can be imputed to microscopic effects inducing an excess of correlations. The best power law behaviour was fitted for larger separations, $|{\bf n}|\geq 30$. It is represented on the graph by the gray line. 

\item $\bra \varep^L \varep^L \ket$, \fref{fig:graphEE}: To exclude microscopy the power law was performed on the subset $|{\bf n}| \geq 10$. After subtraction of $\bra \varep^L \ket^2$, the fit is performed by $\chi^2$ minimization on $x \to ax^b$. 

\item $\bra \varep^L \si^L \si^L \ket$, \fref{fig:graphESS}: A power law fit for insertions $|{\bf n}| \in [8,17]$ returns exponent and offset estimates agreeing with the bootstrap predictions. In particular the fitted structure constant $1.0504\,(195)$ has a striking agreement. The systematic error from the uncertainties of $N_{\si}$ and $N_{\varep}$ are negligible compared to the fit uncertainties and have been omitted.

For a numerical estimation with smaller uncertainty, one option is to remove $\Delta_{\si}$ and $\Delta_{\varep}$ from the fit parameters by replacing them by their bootstrap estimations. Such fit with a single degree of freedom - and restricted to $|{\bf n}| \geq 10$ - returns $C_{\varep \si \si} = 1.05334\,(203)$; in very good agreement with previous estimations.

\item $\bra \varep^L \varep^L \varep^L \ket$, \fref{fig:graphEEE}: Fitting on a power law and reading the exponent gives a close estimate of the energy density scaling weight at $1.4129\,(63)$. The offset fit offers a $C_{\varep \varep \varep}$ numerical estimation of $1.675\,(51)$. This fitting uncertainty is rather large, but the estimate is in a 3$\si$-range of the CFT bootstrap value.

If restricting the fit to a single degree of freedom, fitting on $x \to \alpha x^{-3\Delta_{\varep}}$, our estimation becomes $1.578\,(27)$. The latter fit is in much better agreement with the expected value. This data shows more noise than the three previous ones even though using many more boundary states and thus significantly more measurement time. Indeed, here the decay of the signal is much faster with increasing separation and the noise dominates for separations $> 15$.

\ei

\paragraph{Discussion.}

The general conclusion is that the averages show a behaviour close to their expectation on the infinite plane. Especially the measurements of the $\bra \si^L_{\bf i} \si^L_{\bf j} \ket$ and $\bra \varep^L_{\bf i} \varep^L_{\bf j} \ket_{\rm connected}$ show clear power law profiles typical of planar CFTs. On top of that, the fitted non-universal and universal observables match with previous MonteCarlo studies \citep{ising3dperturbedFiniteSize1,ising3dperturbedFiniteSize2} - which use an entirely different method based on a massive deformation of the CFT to reduce their finite size and boundary effects - and agree within smaller uncertainties with the more precise bootstrap methods \citep{kos2016precision}, see tables \pref{tab:weights} and \pref{tab:structureConstants}.

Further work could include the study of the complete profile of $\bra \varep^L_{\bf i} \si^L_{\bf j} \si^L_{\bf k} \ket_{\rm connected}$ as its CFT formula depends on the assumption that conformal invariance follows from scaling invariance in 3d. The investigation of four point functions such as $\bra \si^L_{\bf i}\si^L_{\bf j} \si^L_{\bf k}\si^L_{\bf l}\ket$ could also bring numerical insights on its expansion on conformal blocks \citep{simmons2016tasi}.

\twocolumn[{%
 \begin{@twocolumnfalse}
\paragraph{Non universal}
\begin{equation}
\label{tab:offsets}
\begin{array}{c|c|c}
\bra \varep^L \ket  & N_{\si} & N_{\varep} \\
\hline \hline
& & \\
.3302047\,(88)& .55247\,(13) & .23068\,(382)
\end{array}
\end{equation}

\paragraph{Universal}
\begin{equation}
\label{tab:weights}
\begin{array}{c|c}
\Delta_{\si} & \Delta_{\varep}\\
\hline \hline
& \\
.518360\,(63)\ \tchoc{.5181489}\footnotemark& 1.41008\,(670)\ \tchoc{1.412625}\\ 
\end{array}
\end{equation}

\paragraph{Structure constants}
\begin{equation}
\begin{array}{c||c}
C_{\varep \si \si} & C_{\varep \varep \varep}\\
\hline \hline
& \\
1.0504\,(195)\ \tchoc{1.0518537\,(41)}& 1.675\,(51)\ \tchoc{1.532435\,(19)}\\
1.05334\,(203) \quad{\text{one d.o.f. fit}}  & 1.578\,(27) \quad{\text{one d.o.f. fit}}
\end{array} 
\label{tab:structureConstants}
\end{equation}
\vspace*{1cm}
\end{@twocolumnfalse}
}]

\footnotetext{Bold chocolate-color numbers are the numerical bootstrap estimations taken from \citep{kos2016precision}}

%


\paragraph{Acknowledgement}

The author would like to express gratitude to B. Doyon for his guidance and numerous advice. The author is funded by a Graduate Teaching Assistantship from KCL Department of Mathematics.

\small{
\bibliography{draftIsing.bib}{}
\bibliographystyle{unsrt}
}
\end{document}